# Tuning hyperparameters of doublet-detection methods for single-cell RNA sequencing data

Nan Miles Xi [a, *] and Angelos Vasilopoulos [a]

[a] *Department of Mathematics and Statistics, Loyola University Chicago, Chicago, IL 60660, USA*

[*] Correspondence: Nan Miles Xi (mxi1@luc.edu)

**Abstract**

The existence of doublets in single-cell RNA sequencing (scRNA-seq) data poses a great challenge in downstream data analysis. Computational doublet-detection methods have been developed to remove doublets from scRNA-seq data. Yet, the default hyperparameter settings of those methods may not provide optimal performance. Here, we propose a strategy to tune hyperparameters for a cutting-edge doublet-detection method. We utilize a full factorial design to explore the relationship between hyperparameters and detection accuracy on 16 real scRNA-seq datasets. The optimal hyperparameters are obtained by a response surface model and convex optimization. We show that the optimal hyperparameters provide top performance across scRNA-seq datasets under various biological conditions. Our tuning strategy can be applied to other computational doublet-detection methods. It also offers insights into hyperparameter tuning for broader computational methods in scRNA-seq data analysis.

# Introduction

Single-cell RNA sequencing (scRNA-seq) is a cutting-edge sequencing technology that can quantify genome-wide gene expression levels in a large number of cells [1,2]. Since its debut, scRNA-seq has been widely applied in various fields, including precision medicine [3], drug discovery [4], cancer therapy [5], and vaccine development [6]. The successful application of scRNA-seq relies on separating and labeling mRNA molecules from different cells. However, results may be confounded by the formation of doublets — when two cells are captured in one reaction volume by chance [7]. Because doublets appear as but are not real cells, they potentially bias downstream scRNA-seq data analysis. For example, doublets may be falsely identified as new cell types in cell clustering analysis [8]. To tackle this issue, the scRNA-seq community has developed computational methods to detect doublets from scRNA-seq data [7–11]. These methods utilize statistical and machine learning models, each with a set of default hyperparameters. Despite rapid development in methodology, one critical question remains untouched: whether default hyperparameters offer the best doublet-detection performance, especially for scRNA-seq datasets generated under various biological conditions.

Here, we systematically explore the optimal hyperparameters of scDblFinder [9], one cutting-edge computational doublet-detection method. We collect detection accuracy data for various hyperparameter combinations from 16 real scRNA-seq datasets with experimentally annotated doublets. Then, we fit a second-degree polynomial regression model with first-order, second-order, and interaction terms of three key hyperparameters. Convex optimization is performed to find the hyperparameters that maximize average detection accuracy across 16 datasets. We show that our optimal hyperparameters significantly improve doublet-detection accuracy over the method's default settings on most datasets. The detection accuracy of our optimal hyperparameters also ranks close to the best performance obtained by exhaustive searches in many datasets. We also apply our tuning strategy to scRNA-seq datasets under various biological conditions using different double-detection measurements. We find similar benefits from hyperparameter tuning, and the optimal hyperparameters vary depending on the biological conditions and accuracy measurements. Our exploratory strategy can be easily extended to other computational doublet-detection methods and provides hyperparameter tuning insights for broader computational methods in scRNA-seq data analysis.

## Datasets

In this study, we utilize 16 public scRNA-seq datasets collected in a previous benchmark study [12]. All datasets contain ground-truth doublet labels annotated by experimental techniques. They are so far the most comprehensive scRNA-seq data collection with ground-truth doublet labels. The datasets cover a wide range of cell types, droplet numbers, gene numbers, and doublet rates, representing various difficulty levels in detecting doublets from scRNA-seq data. Table 1 summarizes the key characteristics of the 16 datasets used in this study. In scRNA-seq experiments, droplets refer to the reaction volumes that encapsulate the cell suspension. While most droplets contain one cell (singlets) as expected, others accidentally encapsulate two cells (doublets). Therefore, we will use "droplet" instead of "cell" to denote one data point in scRNA-seq datasets in the following text.

## Hyperparameter setting

We choose scDblFinder as the target method for exploring optimal hyperparameter settings. The design of scDblFinder can be summarized in the following steps. First, it generates artificial doublets by combining gene expression profiles of two randomly selected droplets in the dataset. Second, it conducts PCA dimension reduction on the union of artificial doublets and true droplets using top highly variable genes. Third, scDblFinder constructs a nearest neighbor network on top of the low-dimensional representations from the dimension reduction. Fourth, it sets different neighborhood sizes to create multiple predictors that will be used for binary classification. Finally, it performs cross-validation to assign a doublet score to each droplet. In each iteration of cross-validation, it trains a gradient boosting model to distinguish true droplets and artificial doublets in the training set, and then assigns each droplet in the test set a doublet probability (doublet score). The design of scDblFinder helps to reduce the impact of batch effects on doublet detection: since cross-validation randomly assigns droplets from different batches to training and test sets, the batch effects will not cause the domain shift problem [13] in the final classification step.

scDblFinder has shown superior performance in previous benchmark studies [12,14]. We consider three key hyperparameters of scDblFinder, i.e., the number of top features, the number of top principal components, and the maximum depths of decision trees. We refer to them as $nf$, $pc$, and

$depth$ moving forward, respectively. These three hyperparameters are discrete numerical variables, and we set each of them to five different levels (Table 2). Therefore, there are $5 \times 5 \times 5 = 125$ hyperparameter combinations in total. In experimental design literature, this is a $3^5$ full factorial design [15]. It allows investigation of the effects of individual hyperparameters, as well as the effects of interactions between different hyperparameters on the performance of doublet detection.

We choose the five values for each hyperparameter according to the following rule. First, we start with the default values of each hyperparameter (Table 2). In general, the default values are selected by the developers based on extensive numerical experiments and thus are likely close to a local optimum in the hyperparameter space. Second, with the default value as the center, we increase or decrease each hyperparameter by one or several fixed step sizes, generating four extra alternative values. We determine the step size and the search space boundaries for each hyperparameter based on the common practice in scRNA-seq data analysis.

The hyperparameter $nf$ refers to the number of highly variable genes used in the downstream analysis. Its value is often set from several hundred to several thousand in many applications. For example, scDblFinder uses 1000 as the default value and Seurat [16], a popular R package for scRNA-seq data analysis, chooses 2000. After including these two values in our search space, we insert 1500, the median value between 1000 and 2000, as the third search value, resulting in a step size of 500. We further expand the search space downward and upward by one step size separately. The final search space for $nf$ is 500, 1000, 1500, 2000, and 2500.

The hyperparameter $pc$ is the number of principal components used in the downstream analysis after performing PCA dimension reduction on highly variable genes. Its value is often set from single digits to several dozen in practice. For example, the Seurat tutorial suggests exploring between 5 to 20 for various scenarios. We start with the default value of 10 and include 5 (the lower bound suggested by Seurat) in the search space, using a step size of 5. We further expand the search space by three step sizes up to 25. The final search space for $pc$ is 5, 10, 15, 20, and 25.

The hyperparameter $depth$ is the maximum depth of decision trees in the gradient boosting model used in scDblFinder. The larger values indicate more complex gradient boosting models in binary classification (singlet vs. doublet). This hyperparameter is often set to below ten in ordinary classification tasks to avoid overfitting. For example, XGBoost [17], a generic gradient boosting package, chooses 6 as the default value. We use scDblFinder's default value of 4 as the center of the search space. With 6 as the maximum and 1 as the step size, we create a final search space for $depth$ including 2, 3, 4, 5, and 6.

## Doublet detection

We use the R package DoubletCollection [14] to execute scDblFinder on 16 real datasets with the 125 hyperparameter combinations listed in Table 2. Since doublet detection is essentially a binary classification task, we use the area under the precision-recall curve (AUPRC) to measure the overall doublet-detection accuracy. After execution, each dataset results in a $125 \times 4$ data matrix, in which the first three columns are $nf$, $pc$, and $depth$, and the last column is AUPRC. Each row in the data matrix represents one combination of three hyperparameters and corresponding AUPRC. The 16 scRNA-seq datasets generate 16 such data matrices. Finally, we merge the 16 data matrices by averaging their AUPRCs for each hyperparameter combination. The final data matrix is $125 \times 4$, which contains the relationship between hyperparameters and overall doublet-detection accuracy. We refer to this data matrix as detection accuracy data moving forward.

## Model setup and optimization

We build a second-degree polynomial regression model to examine the relationship between hyperparameters and doublet-detection accuracy. Specifically, we set average AUPRC as the response variable and the first order of the three hyperparameters, the second order of the three hyperparameters, and their interactions as the independent variables. Model (1) shows the complete setup of this second-degree polynomial regression.

$$AUPRC = \beta_0 + \beta_1 nf + \beta_2 pc + \beta_3 depth + \beta_4 nf^2 + \beta_5 pc^2 + \beta_6 depth^2 + \beta_7 nfpc + \beta_8 nfdepth + \beta_9 pcdepth + \epsilon \quad (1)$$

where $\beta_i, i = 0, 1, \ldots, 9$, are the unknown model parameters, and $\epsilon$ is the random error.

Second-degree polynomial regression is one classic model in the response surface methodology (RSM) [18]. It is commonly used to explore the relationship between several independent variables (hyperparameters) and one response variable (AUPRC) based on a full factorial design [19]. It can obtain an optimal response by estimating hyperparameters' main and quadratic effects and interactions between them. The second-degree polynomial regression balances model complexity and interpretation, while higher-degree models may cause overfitting and are harder to interpret.

We fit this model by least square estimation using detection accuracy data. We perform a $t$-test to assess the significance of estimated parameters $\hat{\beta}_i$ and set 0.01 as the $p$-value cutoff. We find that the first and second orders of $nf$ and $pc$ are significant. Equation (2) shows the estimated model (1) with significant independent variables (including the intercept).

$$AUPRC = 5.444 \times 10^{-1} + 1.016 \times 10^{-5} nf + 1.336 \times 10^{-3} pc - 3.760 \times 10^{-9} nf^2 - 3.484 \times 10^{-5} pc^2 \quad (2)$$

To obtain the $nf$ and $pc$ that maximize AUPRC, we take the partial derivative of AUPRC in respect of $nf$ and $pc$ in (2) and let the derivatives equal zero simultaneously.

$$\begin{cases} \dfrac{\partial AUPRC}{\partial nf} = 1.016 \times 10^{-5} - 7.520 \times 10^{-9} nf = 0 \\ \dfrac{\partial AUPRC}{\partial pc} = 1.336 \times 10^{-3} - 6.968 \times 10^{-5} pc = 0 \end{cases} \quad (3)$$

Solving (3) gives the optimal $nf$ and $pc$ (after rounding to the nearest integers).

$$\begin{cases} nf = 1352 \\ pc = 19 \end{cases} \quad (4)$$

**Model diagnostics**

The 16 scRNA-seq datasets are generated by different sequencing protocols using various doublet annotation techniques. The error terms in model (1) may have non-constant variance, causing the heterogeneity issue. We conduct model diagnostics to examine the existence and severity of heterogeneity. First, we plot the residue against the fitted value of model (1). Supplementary Figure S1a shows that most residues have constant variance with no obvious patterns. The only concern is on the left, where the six residues may have a smaller variance. Second, we perform a Breusch-Pagan test [20] for heterogeneity. With the $p$-value as 0.451, we fail to reject the null hypothesis that constant variance is present.

Additionally, we perform a sensitivity analysis to examine the robustness of model (1). We conduct a natural log transformation and a square root transformation on the response variable AUPRC in the detection accuracy data. We then fit model (1) on the two transformed datasets and obtain the optimal hyperparameters using the same optimization method in (3). We find that the significant hyperparameters in model (1) and their optimal values are identical (after rounding to the nearest integers) to the results without transformation (Supplementary Table S1). The patterns of residue plots (Supplementary Figures S1b and S1c) are also similar to those before transformation (Supplementary Figure S1a). Log transformation and square root transformation on the response variable are common remedies for heterogeneity. If heterogeneity exists, then these two transformations would significantly change the model fitting, optimization, and residue plot. Similar results before and after transformations indicate that the heterogeneity is very mild, if any. We suspect that the heterogeneity is largely reduced or removed by averaging the AUPRCs of 16 datasets when creating the detection accuracy data (on which we fit model (1)).

## Optimal hyperparameter evaluation

The optimal hyperparameters in equation (4) are obtained by maximizing the *average* AUPRC across 16 scRNA-seq datasets. To examine if these parameters can improve doublet-detection accuracy on individual datasets, we execute scDblFinder with $nf$ and $pc$ as in (4) on all 16 scRNA-seq datasets. Since hyperparameter $depth$ is not significant in model (1), we set it to the default value in the execution. Table 3 compares the AUPRCs of the optimal hyperparameters in equation (4), the method's default hyperparameters, and the maximal AUPRCs achieved by one of 125 hyperparameter combinations. Our optimal hyperparameters outperform the method's performance with default settings on 12 out of 16 scRNA-seq datasets. Figure 1 summarizes the AUPRC improvement by hyperparameters tuning over the method's default settings. The most significant improvement is over 5% on dataset pbmc-1B-dm. There are eight datasets on which the improvement is over 2%. Figure 2 shows each dataset's AUPRC ranking under optimal hyperparameters among 125 hyperparameter combinations. We can see that the AUPRCs of optimal hyperparameters rank at or higher than the 20th percentile on ten datasets. The highest ranking is 3rd on dataset pdx-MULTI. The optimal hyperparameters also achieve the 50th percentile or higher on all 16 datasets.

## Tuning hyperparameters for various biological conditions

The previous analysis presents the optimal hyperparameters based on the average of 16 scRNA-seq datasets. In practice, users mainly conduct doublet detection on datasets generated under specific biological or technical conditions. Those datasets need unique hyperparameter settings to achieve optimal performance. To demonstrate the generability of our tuning strategy to those applications, we replicate the hyperparameter optimization on two subsets of 16 datasets.

The first is the pbmc-related subset, including six datasets: pbmc-1A-dm, pbmc-1B-dm, pbmc-1C-dm, pbmc-2ctrl-dm, pbmc-2stim-dm, and pbmc-ch. We find that the optimal $pc$ is 18 (after rounding to the nearest integer), with the other two hyperparameters insignificant (Supplementary Table S2). Supplementary Table S3 compares the AUPRCs of the optimal hyperparameters, the method's default hyperparameters, and the maximal AUPRCs achieved by one of 125 hyperparameter combinations on pbmc-related datasets. Supplementary Figure S2a shows each dataset's AUPRC improvement by hyperparameters optimization over the method's default settings, and AUPRC ranking under optimal hyperparameters among 125 hyperparameter combinations. Compared with the optimization on all 16 datasets (Figures 1 and 2), the AUPRC improvement is greater with hyperparameters specifically tuned for pbmc-related datasets.

The second subset includes the three HMEC-related datasets: HMEC-orig-MULTI, HMEC-rep-MULTI, and HEK-HMEC-MULTI. We find that the optimal $nf$ is 1520 (after rounding to the nearest integer), with the other two hyperparameters insignificant (Supplementary Table S2). Supplementary Table S4 compares the AUPRCs of the optimal hyperparameters, the method's default hyperparameters, and the maximal AUPRCs achieved by one of 125 hyperparameter combinations on HMEC-related datasets. Supplementary Figure S2b shows each dataset's

AUPRC improvement by hyperparameters optimization over the method's default settings, and AUPRC ranking under optimal hyperparameters among 125 hyperparameter combinations. Compared with the optimization on all 16 datasets (Figures 1 and 2), the AUPRC improvement is greater with hyperparameters specifically tuned for HMEC-related datasets.

The two analyses provide guidance for choosing appropriate hyperparameters for specific biological conditions. Future studies can easily expand our tuning strategies to other cell types or platforms if datasets with doublet labels under more diverse biological and technical conditions are available. For example, users can optimize hyperparameters for different sequencing protocols (Smart-seq2, Drop-seq, Chromium, et al.) or the combinations of biological and technical conditions (pbmc and Drop-seq, HMEC and Smart-seq2, et al.).

## Tuning hyperparameters for various measurements

The previous analyses use AUPRC, an overall accuracy measurement of doublet detection, as the optimization objective. In practice, users may also be interested in the method's capacity to identify doublets or singlets, i.e., the true positive or negative rate. Different from AUPRC, the calculation of true positive/negative rate requires a user-specified doublet rate. The true doublet rate is typically unknown to the users and needs to be estimated based on the sequencing platform, sequencing throughput, and prior knowledge [12,21]. Because optimization relies on the doublet rate, it is infeasible to find universal optimal hyperparameters for the true positive/negative rate.

To provide hyperparameter guidance under this scenario, we set the doublet rates to their true values for each dataset (Table 1) and calculate the corresponding true positive/negative rates for 125 hyperparameter combinations. Then we replicate our tuning strategy using these two measurements as objectives. We find that the optimal maximum depths of decision trees ($depth$) is 5 for both measurements (after rounding to the nearest integer), with the other two hyperparameters insignificant (Supplementary Table S2). Supplementary Figure S3a and Table S5 show each dataset's true positive rate improvement by hyperparameters optimization over the method's default settings and true positive rate ranking under optimal hyperparameters among 125 hyperparameter combinations. Most datasets exhibit similar improvement as AUPRC, except for J293t-dm, with a 29% increase, significantly larger than others. Such a difference indicates this dataset's unique biological and technical characteristics, which require stronger hyperparameter tuning efforts.

Supplementary Figure S3b and Table S6 show each dataset's true negative rate improvement and ranking. Although most datasets still benefit from hyperparameter tuning, the improvement of the true negative rate is milder (below 1%) compared to other metrics. One reason is that the true negative rates under default hyperparameters are already high on many datasets (above 0.95, Supplementary Table S6), limiting the improvement space by hyperparameter optimization. It is worth noting that the optimal hyperparameters and corresponding true positive/negative rates are obtained using the true doublet rates. If users choose different doublet rates, the optimization results will be different. It is straightforward to generalize our tuning strategy in those cases.

## Discussion

The existence of doublets is a key confounder in scRNA-seq data analysis. With the wide application of scRNA-seq technologies, much effort has been invested in developing computational doublet-detection methods. Such methods are primarily based on statistical and machine learning algorithms and are sensitive to hyperparameter configurations [22]. Although most methods provide a set of default hyperparameters, they cannot guarantee the best doublet-detection performance universally, especially when scRNA-seq data are generated under various biological conditions [23,24].

In this study, we utilize a full factorial design to build a model of hyperparameters and overall doublet-detection accuracy based on a leading method, scDblFinder, and 16 scRNA-seq datasets. The optimal hyperparameter combination obtained by convex optimization not only surpasses the default setting but also offers close-to-best detection accuracy on many datasets. We expand our optimization strategy to two subgroups of 16 datasets separately, providing optimal parameters for various biological conditions. We show that our method can also be applied to optimize different measurements of doublet-detection accuracy.

The improved doublet-detection performance by hyperparameter tuning presents several insights regarding the data generalization and doublet annotation mechanisms. First, there are two datasets, hm-6k and hm-12k, whose doublets are annotated by species mixture [25]. Both have lower AUPRCs using optimal hyperparameters than default hyperparameters (Figure 1). In contrast, most datasets generated by the other three doublet annotation techniques, i.e., cell hashing [26], demuxlet [27], and MULTI-seq [28], benefit from hyperparameter tuning. One possible reason is due to their different doublet-annotation mechanisms. While species mixture can only annotate doublets from two species, the other three techniques utilize oligo-tagged antibody, SNP, or lipid-tagged index to label doublets from much broader sources. Consequently, the true doublets in hm-6k and hm-12k are likely undercounted, causing their inconsistent optimization results.

Second, the hyperparameter tuning fails to improve the AUPRC for dataset pbmc-ch, even if the hyperparameters are optimized specifically for pbmc-related datasets (Supplementary Figure S2). In contrast, optimal hyperparameters consistently improve AUPRC for the other five pbmc-related datasets, and the improvements are greater with specifically tuned hyperparameters (Figure 1 and Supplementary Figure S2). Such discrepancy is potentially due to the different doublet annotation techniques (cell hashing vs. demuxlet) and batch effects among those datasets. Further investigations, especially from the experimental perspective, are needed to reveal the impacts of these two factors on doublet detection.

Third, the optimal hyperparameters vary depending on the biological conditions and optimization objectives. There are no universal hyperparameters adaptive to all scenarios. The significant hyperparameters when optimizing AUPRC across all 16 datasets are $pc$ and $nf$, with optimal values as 19 and 1252, respectively (Supplementary Table S2). If optimized on pbmc-related datasets, $nf$ is no longer significant, and the optimal value of $pc$ changes to 18. If optimized on HMEC-related datasets, $pc$ is no longer significant, and the optimal value of $nf$ changes to 1520. Depth is the only significant hyperparameter when optimizing the true positive and negative rate on all 16 datasets, with optimal values as 5 in both cases. This result indicates that existing and future doublet-detection methods need to fine-tune hyperparameters for a variety of biological conditions and accuracy measurements.

In summary, doublet detection is one essential step in the quality control of scRNA-seq data analysis. The hyperparameter configuration significantly impacts the performance of computational doublet-detection methods. Our study is the first attempt to systematically explore the optimal hyperparameters under various biological conditions and optimization objectives. Our study provides much-needed guidance for hyperparameter tuning in computational doublet-detection methods. Future directions of our study include increasing the exploratory space by utilizing advanced experimental design strategies, e.g., space-filling design [29,30] and fractional factorial design [31,32]. Another direction is to optimize hyperparameters for other doublet-detection methods by our tuning strategy. More scRNA-seq datasets with experimentally annotated doublets could also be incorporated into the tuning process to enhance the generalizability of optimal hyperparameters.

## Data Availability

The 16 scRNA-seq datasets used in this study are available at Zenodo repository
https://doi.org/10.5281/zenodo.4562782

# References

1. Kolodziejczyk, A. A., Kim, J. K., Svensson, V., Marioni, J. C. & Teichmann, S. A. The technology and biology of single-cell RNA sequencing. *Mol. Cell* **58**, 610–620 (2015).

2. Saliba, A.-E., Westermann, A. J., Gorski, S. A. & Vogel, J. Single-cell RNA-seq: advances and future challenges. *Nucleic Acids Research* vol. 42 8845–8860 Preprint at https://doi.org/10.1093/nar/gku555 (2014).

3. Wiedmeier, J. E., Noel, P., Lin, W., Von Hoff, D. D. & Han, H. Single-Cell Sequencing in Precision Medicine. *Cancer Treat. Res.* **178**, 237–252 (2019).

4. Aissa, A. F. *et al.* Single-cell transcriptional changes associated with drug tolerance and response to combination therapies in cancer. *Nat. Commun.* **12**, 1–25 (2021).

5. Sun, G. *et al.* Single-cell RNA sequencing in cancer: Applications, advances, and emerging challenges. *Molecular Therapy - Oncolytics* **21**, 183–206 (2021).

6. Noé, A., Cargill, T. N., Nielsen, C. M., Russell, A. J. C. & Barnes, E. The Application of Single-Cell RNA Sequencing in Vaccinology. *J Immunol Res* **2020**, 8624963 (2020).

7. Wolock, S. L., Lopez, R. & Klein, A. M. Scrublet: Computational Identification of Cell Doublets in Single-Cell Transcriptomic Data. *Cell Syst* **8**, 281-291.e9 (2019).

8. McGinnis, C. S., Murrow, L. M. & Gartner, Z. J. DoubletFinder: Doublet Detection in Single-Cell RNA Sequencing Data Using Artificial Nearest Neighbors. *Cell Syst* **8**, 329-337.e4 (2019).

9. Germain, P.-L., Lun, A., Garcia Meixide, C., Macnair, W. & Robinson, M. D. Doublet identification in single-cell sequencing data using scDblFinder. *F1000Res.* **10**, 979 (2021).

10. Bais, A. S. & Kostka, D. scds: computational annotation of doublets in single-cell RNA sequencing data. *Bioinformatics* Preprint at https://doi.org/10.1093/bioinformatics/btz698 (2019).

11. Bernstein, N. J. *et al.* Solo: Doublet Identification in Single-Cell RNA-Seq via Semi-Supervised Deep Learning. *Cell Syst* **11**, 95-101.e5 (2020).

12. Xi, N. M. & Li, J. J. Benchmarking Computational Doublet-Detection Methods for Single-Cell RNA Sequencing Data. *Cell Systems* **12**, 176-194.e6 (2021).

13. Redko, I., Morvant, E., Habrard, A., Sebban, M. & Bennani, Y. *Advances in Domain Adaptation Theory*. (Elsevier, 2019).

14. Xi, N. M. & Li, J. J. Protocol for executing and benchmarking eight computational doublet-detection methods in single-cell RNA sequencing data analysis. *STAR Protocols* **2**, 100699 (2021).

15. Steinberg, D. M. & Hunter, W. G. Experimental Design: Review and Comment. *Technometrics* **26**, 71–97 (1984).

16. Hao, Y. *et al.* Integrated analysis of multimodal single-cell data. *Cell* **184**, 3573-3587.e29 (2021).

17. Chen, T. & Guestrin, C. XGBoost: A Scalable Tree Boosting System. in *Proceedings of the 22nd ACM SIGKDD International Conference on Knowledge Discovery and Data Mining* 785–794 (Association for Computing Machinery, 2016).

18. Box, G. E. P. & Draper, N. R. Empirical model-building and response surfaces. *Wiley series in probability and mathematical statistics.* **669**, (1987).

19. Box, G. E. P. & Wilson, K. B. On the Experimental Attainment of Optimum Conditions. *J. R. Stat. Soc. Series B Stat. Methodol.* **13**, 1–45 (1951).

20. Breusch, T. & Pagan, A. A Simple Test for Heteroscedasticity and Random Coefficient Variation. *Econometrica* **47**, 1287–1294 (1979).

21. Luecken, M. D. & Theis, F. J. Current best practices in single-cell RNA-seq analysis: a tutorial. *Mol. Syst. Biol.* **15**, (2019).

22. Probst, P., Boulesteix, A. L. & Bischl, B. Tunability: Importance of hyperparameters of machine learning algorithms. *J. Mach. Learn. Res.* (2019).

23. Hu, Q. & Greene, C. S. Parameter tuning is a key part of dimensionality reduction via deep variational autoencoders for single cell RNA transcriptomics. *Pac. Symp. Biocomput.* **24**, 362–373 (2019).

24. Raimundo, F., Vallot, C. & Vert, J.-P. Tuning parameters of dimensionality reduction methods for single-cell RNA-seq analysis. *Genome Biol.* **21**, 212 (2020).

25. Alles, J. *et al.* Cell fixation and preservation for droplet-based single-cell transcriptomics. *BMC Biol.* **15**, (2017).

26. Stoeckius, M. *et al.* Cell Hashing with barcoded antibodies enables multiplexing and doublet detection for single cell genomics. *Genome Biol.* **19**, 224 (2018).

27. Kang, H. M. *et al.* Multiplexed droplet single-cell RNA-sequencing using natural genetic variation. *Nat. Biotechnol.* **36**, 89–94 (2018).

28. McGinnis, C. S. *et al.* MULTI-seq: sample multiplexing for single-cell RNA sequencing using lipid-tagged indices. *Nature Methods* vol. 16 619–626 Preprint at https://doi.org/10.1038/s41592-019-0433-8 (2019).

29. Wang, L., Xiao, Q. & Xu, H. Optimal maximin -distance Latin hypercube designs based on good lattice point designs. *Ann. Stat.* (2018).

30. Wang, L., Sun, F., Lin, D. K. J. & Liu, M.-Q. CONSTRUCTION OF ORTHOGONAL SYMMETRIC LATIN HYPERCUBE DESIGNS. *Stat. Sin.* **28**, 1503–1520 (2018).

31. Wang, L., Xu, H. & Liu, M.-Q. Fractional factorial designs for Fourier-cosine models. *Metrika* (2022) doi:10.1007/s00184-022-00881-2.

32. Wang, L. & Xu, H. A class of multilevel nonregular designs for studying quantitative factors. *Stat. Sin.* (2022) doi:10.5705/ss.202020.0223.

# Figures

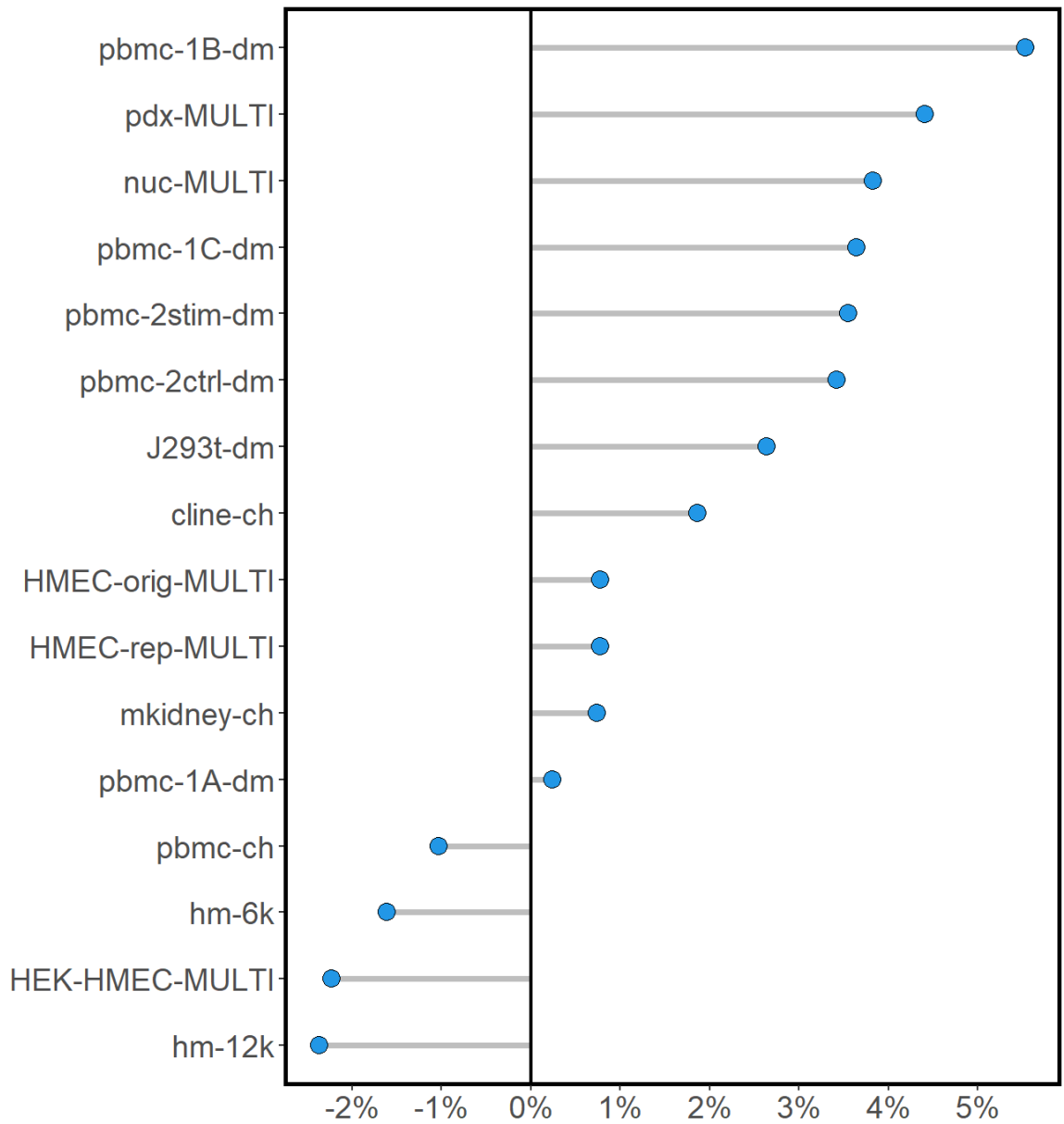

**Figure 1.** AUPRC improvement by hyperparameter optimization over the method's default settings on 16 scRNA-seq datasets.

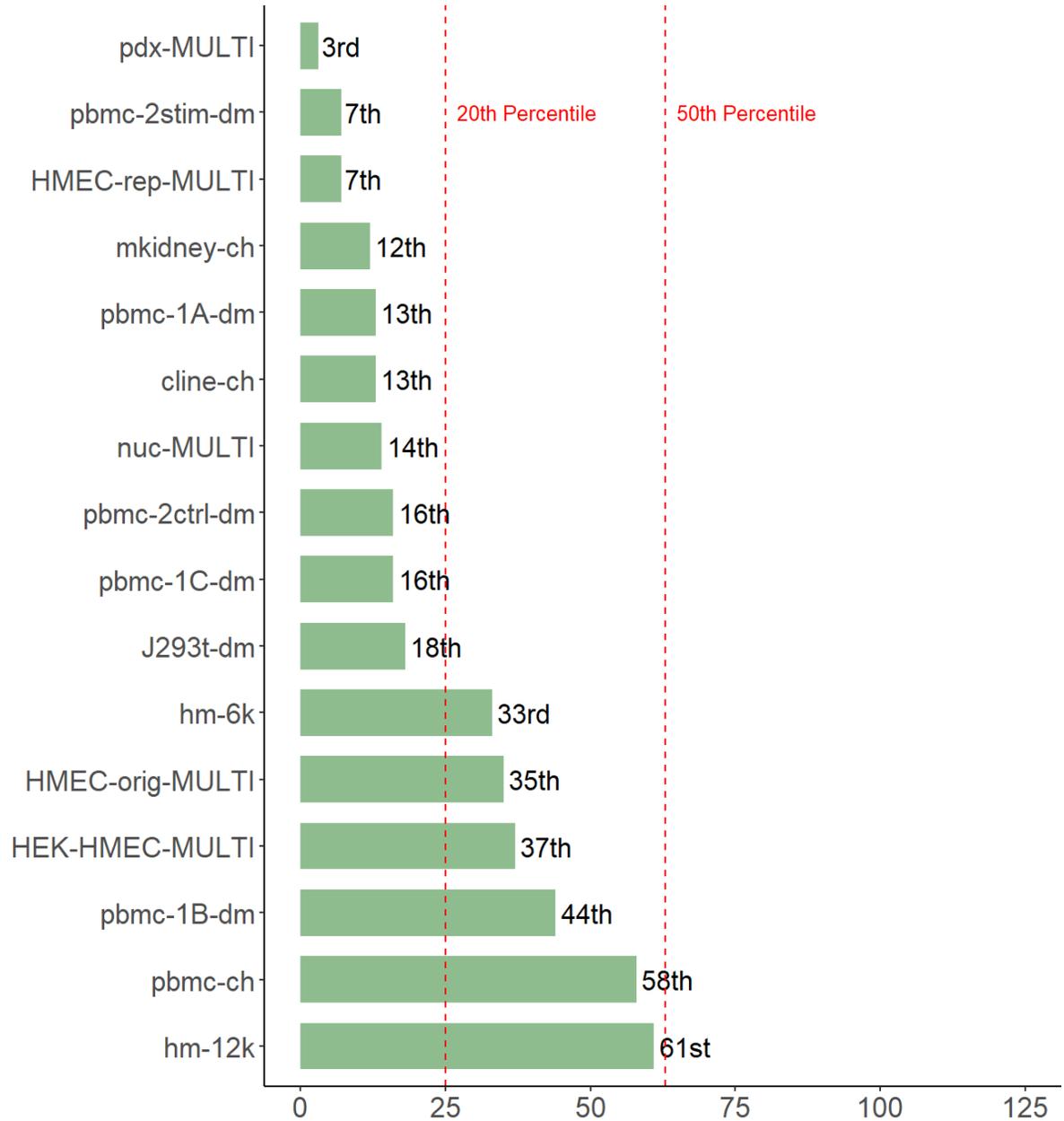

**Figure 2.** Rankings of optimal hyperparameter performances by AUPRC among 125 hyperparameter combinations on 16 datasets.

## Tables

Table 1. The 16 scRNA-seq datasets with experimentally annotated doublets in this study.

| Dataset | Cell type | Droplet # | Gene # | Doublet rate | Doublet annotation technique |
|---|---|---|---|---|---|
| pbmc-ch | pbmc | 15272 | 21639 | 16.66% | Cell hashing [26] |
| cline-ch | HEK293T, K562, KG1, THP1 | 7954 | 25221 | 18.42% | Cell hashing |
| mkidney-ch | mouse kidney | 21179 | 18940 | 37.31% | Cell hashing |
| hm-12k | HEK293T, NIH3T3 | 12820 | 15106 | 5.69% | Species mixture [25] |
| hm-6k | HEK293T, NIH3T3 | 6806 | 15080 | 2.51% | Species mixture |
| pbmc-1A-dm | pbmc | 3298 | 15170 | 3.64% | Demuxlet [27] |
| pbmc-1B-dm | pbmc | 3790 | 15143 | 3.43% | Demuxlet |
| pbmc-1C-dm | pbmc | 5270 | 15865 | 6.00% | Demuxlet |
| pbmc-2ctrl-dm | pbmc | 13913 | 17584 | 11.49% | Demuxlet |
| pbmc-2stim-dm | pbmc | 13916 | 17315 | 11.72% | Demuxlet |
| J293t-dm | jurkat, HEK293T | 500 | 16374 | 8.40% | Demuxlet |
| pdx-MULTI | human breast cancer, mouse immune | 10296 | 14025 | 12.79% | MULTI-seq [28] |
| HMEC-orig-MULTI | HMEC | 26426 | 24199 | 13.50% | MULTI-seq |
| HMEC-rep-MULTI | HMEC | 10580 | 17473 | 31.02% | MULTI-seq |
| HEK-HMEC-MULTI | HEK293T, HMEC | 10641 | 23982 | 4.60% | MULTI-seq |
| nuc-MULTI | nuclei (HEK293T, MEF, Jurkat) | 5578 | 21490 | 8.52% | MULTI-seq |

**Table 2.** The three hyperparameters and their default and exploratory values in this study.

| Hyperparameter | Default values | Exploratory values |
| --- | --- | --- |
| $nf$ | 1000 | 500, 1000, 1500, 2000, 2500 |
| $pc$ | 10 | 5, 10, 15, 20, 25 |
| $depth$ | 4 | 2, 3, 4, 5, 6 |

**Table 3.** The AUPRC of doublet detection under optimal and default hyperparameters. The last column shows the highest AUPRC achieved by one of the 125 hyperparameter combinations. The highest AUPRC between the optimum and default of each dataset is underscored.

| Dataset | Optimum | Default | Maximum |
|---|---|---|---|
| cline-ch | <u>0.4280</u> | 0.4202 | 0.4369 |
| HEK-HMEC-MULTI | 0.4723 | <u>0.4830</u> | 0.4966 |
| HEK-orig-MULTI | <u>0.4911</u> | 0.4873 | 0.5054 |
| hm-12k | 0.9281 | <u>0.9506</u> | 0.9850 |
| hm-6k | 0.9737 | <u>0.9896</u> | 0.9972 |
| HMEC-rep-MULTI | <u>0.6010</u> | 0.5964 | 0.6020 |
| J293t-dm | <u>0.2052</u> | 0.1999 | 0.2525 |
| mkidney-ch | <u>0.6125</u> | 0.6080 | 0.6183 |
| nuc-MULTI | <u>0.4600</u> | 0.4430 | 0.4704 |
| pbmc-1A-dm | <u>0.5454</u> | 0.5441 | 0.5693 |
| pbmc-1B-dm | <u>0.4375</u> | 0.4145 | 0.4818 |
| pbmc-1C-dm | <u>0.5953</u> | 0.5744 | 0.6082 |
| pbmc-2ctrl-dm | <u>0.6980</u> | 0.6749 | 0.7088 |
| pbmc-2stim-dm | <u>0.7003</u> | 0.6763 | 0.7124 |
| pbmc-ch | 0.6405 | <u>0.6472</u> | 0.6520 |
| pdx-MULTI | <u>0.4457</u> | 0.4268 | 0.4477 |

# Supplementary Figures

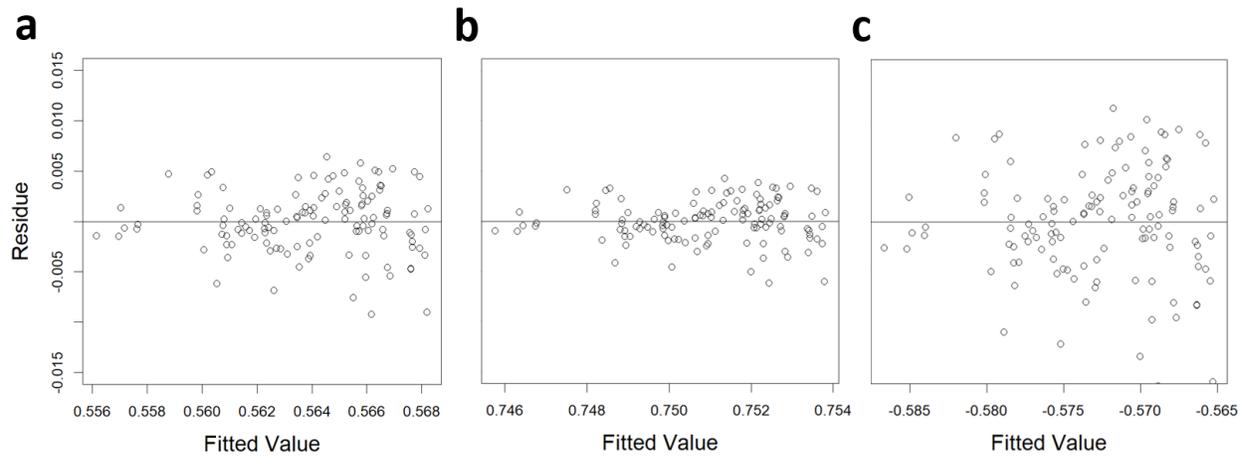

**Supplementary Figure S1**. Residue plots of model (1) on datasets before and after transformation on the response variable. All three figures have the same residue scale. **a**, No transformation. **b**, Square-root-transformation on response variable AUPRC. **c**, Natural-log-transformation on response variable AURPC.

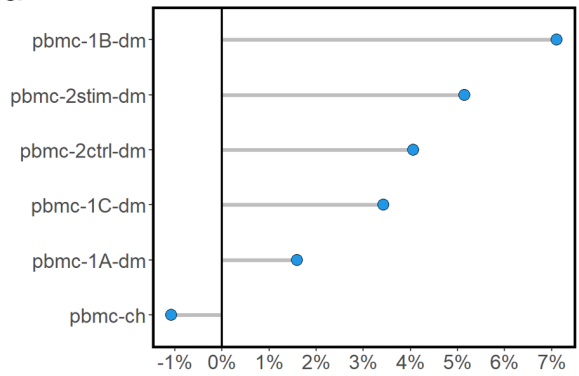
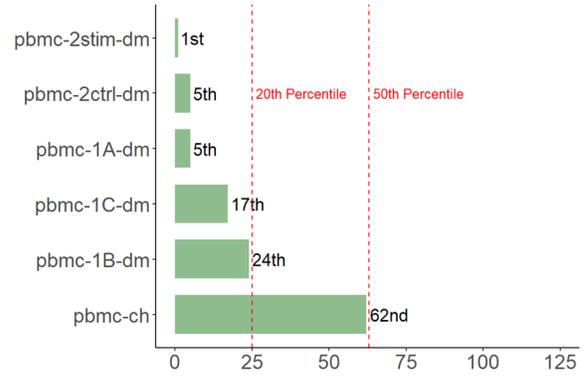
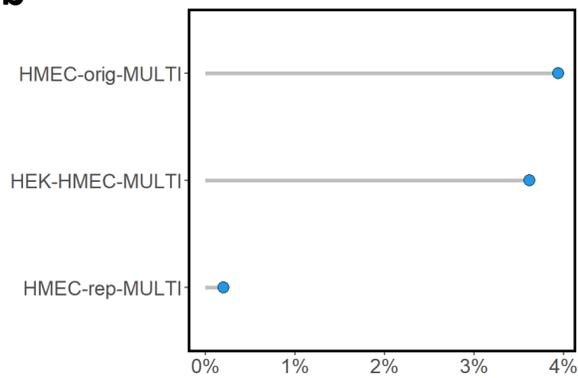
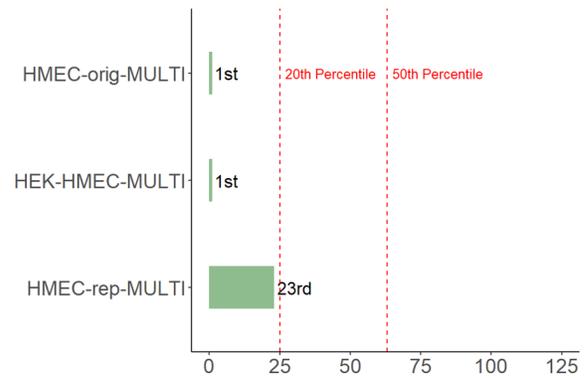

**Supplementary Figure S2**. AUPRC improvement and ranking with hyperparameters optimized on datasets of different biological conditions. **a**, Optimized on pbmc-related datasets. **b**, Optimized on HMEC-related datasets.

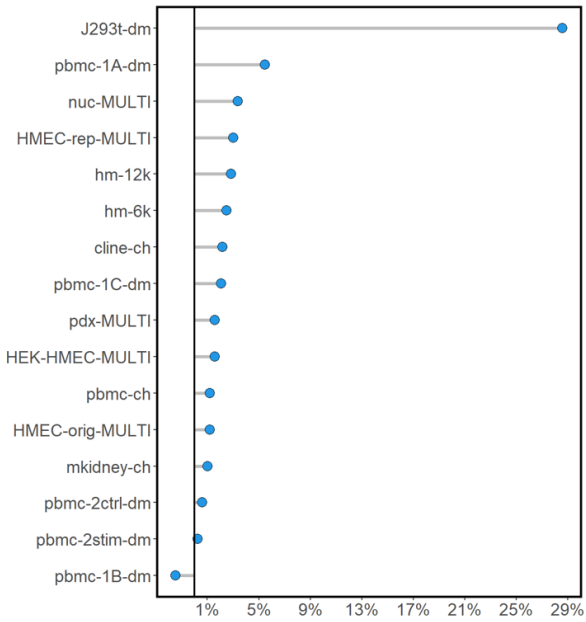
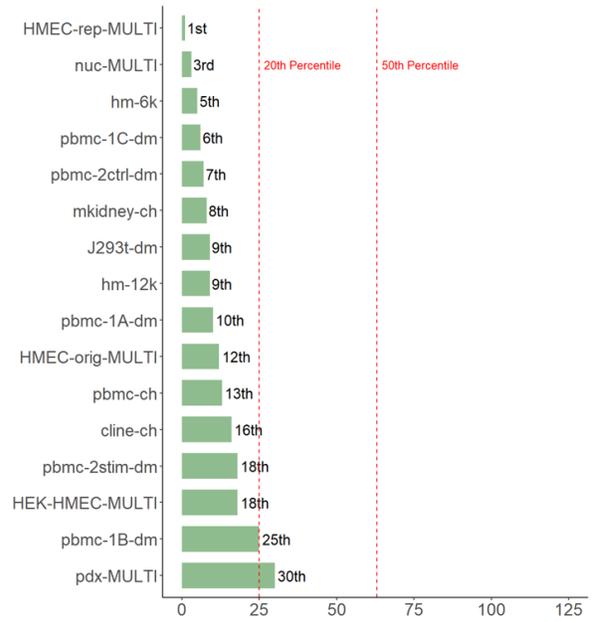
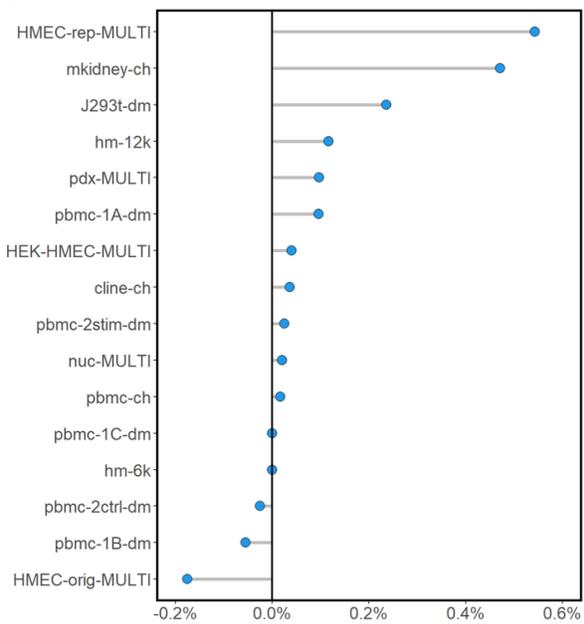
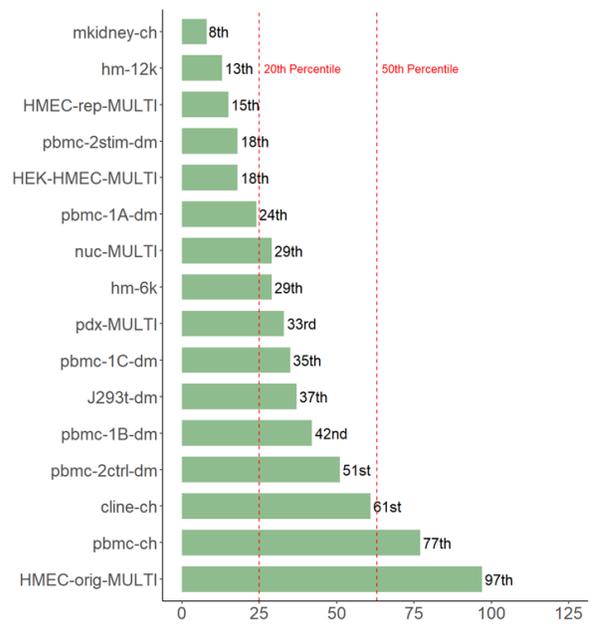

**Supplementary Figure S3**. Double-detection improvement and ranking in terms of different measurements. **a**, Measured by true positive rate. **b**, Measured by true negative rate. The hyperparameters are optimized on all 16 scRNA-seq datasets

# Supplementary Tables

**Supplementary Table S1**. The significant hyperparameters and their raw optimal values before and after transformation on response variable AUPRC.

| Transformation | Significant hyperparameter | Optimal value |
|---|---|---|
| No transformation | $nf, pc$ | nf = 1351.661, pc = 19.181 |
| Square root | $nf, pc$ | nf = 1352.087, pc = 19.189 |
| Natural log | $nf, pc$ | nf = 1352.509, pc = 19.196 |

**Supplementary Table S2.** The significant hyperparameters in model (1) and their optimal values (rounding to the nearest integers) optimized on different biological conditions and accuracy metrics of doublet detection.

| Optimization subset | Optimization metric | Significant hyperparameter | Optimal value |
|---|---|---|---|
| All | AUPRC | $pc, nf$ | $pc = 19$, $nf = 1352$ |
| pbmc-related | AUPRC | $pc$ | $pc = 18$ |
| HMEC-related | AUPRC | $nf$ | $nf = 1520$ |
| All | True positive rate | $depth$ | $depth = 5$ |
| All | True negative rate | $depth$ | $depth = 5$ |

**Supplementary Table S3.** The AUPRC of doublet detection under optimal and default hyperparameters on pbmc-related datasets. The last column shows the highest AUPRC achieved by one of the 125 hyperparameter combinations. The highest AUPRC between the optimum and default of each dataset is underscored.

| Dataset | Optimum | Default | Maximum |
| --- | --- | --- | --- |
| pbmc-1A-dm | <u>0.5528</u> | 0.5441 | 0.5693 |
| pbmc-1B-dm | <u>0.4440</u> | 0.4145 | 0.4818 |
| pbmc-1C-dm | <u>0.5941</u> | 0.5744 | 0.6082 |
| pbmc-2ctrl-dm | <u>0.7023</u> | 0.6749 | 0.7088 |
| pbmc-2stim-dm | <u>0.7111</u> | 0.6763 | 0.7124 |
| pbmc-ch | 0.6402 | <u>0.6472</u> | 0.6520 |

**Supplementary Table S4.** The AUPRC of doublet detection under optimal and default hyperparameters on HMEC-related datasets. The last column shows the highest AUPRC achieved by one of the 125 hyperparameter combinations. The highest AUPRC between the optimum and default of each dataset is underscored.

| Dataset | Optimum | Default | Maximum |
| --- | --- | --- | --- |
| HEK-HMEC-MULTI | <u>0.5005</u> | 0.4830 | 0.4966 |
| HEK-orig-MULTI | <u>0.5065</u> | 0.4873 | 0.5054 |
| HMEC-rep-MULTI | <u>0.5976</u> | 0.5964 | 0.6020 |

**Supplementary Table S5.** The true positive rate of doublet detection under optimal and default hyperparameters. The last column shows the highest true positive rate achieved by one of the 125 hyperparameter combinations. The larger true positive rate between optimum and default in each dataset is underscored.

| Dataset | Optimum | Default | Maximum |
| --- | --- | --- | --- |
| cline-ch | <u>0.3543</u> | 0.3468 | 0.3604 |
| HEK-HMEC-MULTI | <u>0.5297</u> | 0.5215 | 0.5358 |
| HEK-orig-MULTI | <u>0.8973</u> | 0.8726 | 0.9233 |
| hm-12k | <u>0.9591</u> | 0.9357 | 0.9649 |
| hm-6k | <u>0.5028</u> | 0.4969 | 0.5121 |
| HMEC-rep-MULTI | <u>0.5314</u> | 0.5158 | 0.5314 |
| J293t-dm | <u>0.2143</u> | 0.1667 | 0.2381 |
| mkidney-ch | <u>0.5875</u> | 0.5816 | 0.5894 |
| nuc-MULTI | <u>0.4505</u> | 0.4358 | 0.4505 |
| pbmc-1A-dm | <u>0.6417</u> | 0.6083 | 0.6500 |
| pbmc-1B-dm | 0.5154 | <u>0.5231</u> | 0.5462 |
| pbmc-1C-dm | <u>0.6266</u> | 0.6139 | 0.6266 |
| pbmc-2ctrl-dm | <u>0.7196</u> | 0.7153 | 0.7240 |
| pbmc-2stim-dm | <u>0.7192</u> | 0.7174 | 0.7253 |
| pbmc-ch | <u>0.6346</u> | 0.6271 | 0.6369 |
| pdx-MULTI | <u>0.4374</u> | 0.4305 | 0.4487 |

**Supplementary Table S6.** The true negative rate of doublet detection under optimal and default hyperparameters. The last column shows the highest true negative rate achieved by one of the 125 hyperparameter combinations. The larger true negative rate between optimum and default in each dataset is underscored.

| Dataset | Optimum | Default | Maximum |
| --- | --- | --- | --- |
| cline-ch | <u>0.8530</u> | 0.8527 | 0.8558 |
| HEK-HMEC-MULTI | <u>0.9774</u> | 0.9770 | 0.9777 |
| HEK-orig-MULTI | <u>0.9936</u> | 0.9925 | 0.9955 |
| hm-12k | <u>0.9986</u> | 0.9986 | 0.9994 |
| hm-6k | 0.9199 | <u>0.9216</u> | 0.9239 |
| HMEC-rep-MULTI | <u>0.7868</u> | 0.7825 | 0.7895 |
| J293t-dm | <u>0.9279</u> | 0.9258 | 0.9323 |
| mkidney-ch | <u>0.7546</u> | 0.7511 | 0.7558 |
| nuc-MULTI | <u>0.9479</u> | 0.9477 | 0.9490 |
| pbmc-1A-dm | <u>0.9865</u> | 0.9855 | 0.9871 |
| pbmc-1B-dm | 0.9831 | <u>0.9836</u> | 0.9844 |
| pbmc-1C-dm | <u>0.9756</u> | 0.9756 | 0.9764 |
| pbmc-2ctrl-dm | 0.9629 | <u>0.9631</u> | 0.9643 |
| pbmc-2stim-dm | <u>0.9629</u> | 0.9626 | 0.9637 |
| pbmc-ch | <u>0.9257</u> | 0.9256 | 0.9276 |
| pdx-MULTI | <u>0.9176</u> | 0.9167 | 0.9194 |